\documentclass[twocolumn,aps,prl,superscriptaddress,amsmath,amssymb,showpacs]{revtex4}

\usepackage{graphics}
\usepackage{graphicx}
\usepackage{dcolumn}
\usepackage{bm}

\begin{document}

\title{The effect of non-linear quantum electrodynamics on relativistic transparency and laser absorption in ultra-relativistic plasmas}

\author{Peng Zhang}
\author{A. G. R. Thomas}
\affiliation{Department of Nuclear Engineering and Radiological Sciences, \\
University of Michigan, Ann Arbor, MI 48109-2104, USA}

%
%
\author{C. P. Ridgers}
\affiliation{York Plasma Institute,Physics Department, University of York, UK}


\date{\today}

\begin{abstract}
With the aid of large-scale three-dimensional QED-PIC simulations, we describe a realistic experimental configuration to measure collective effects that couple strong field quantum electrodynamics to plasma kinetics. For two counter propagating lasers interacting with a foil at intensities  exceeding $10^{22}$ Wcm$^{-2}$, a binary result occurs; when quantum effects are included, a foil that classically would effectively transmit the laser pulse becomes opaque. This is a dramatic change in plasma  behavior, directly as a consequence of the coupling of radiation reaction and pair production to plasma dynamics.
 
\end{abstract}

\maketitle
When the next generation of 10 PW lasers currently under construction are built\cite{Rus13,Mou07,Mar10,Ger07,Ger10}, there are likely to be a few surprises in the way that they interact with matter. At the extreme intensities expected to be reached in the laser focus ($>10^{22}$ Wcm$^{-2}$) matter is rapidly ionised and the electrons in the resulting plasma are accelerated to such ultra-relativistic energies that the electric field they experience in their rest frame may reach the critical or Schwinger field of quantum electrodynamics (QED), $E_s=1.3\times10^{18}\;{\rm Vm}^{-1}$ \cite{Sch51}. This field is strong enough to  break down the quantum vacuum into electron-positron pairs and is  the threshold at which strong-field QED effects start to play a crucial role \cite{Bel08,Fed10,Bul10,Ner11}. 
  
 The new plasma state that is created is similar to that thought to exist in extreme astrophysical environments including the magnetospheres of pulsars and active black holes \cite{Gol69,Bla77}. Here collective plasma processes are strongly affected by pair creation and radiation reaction \cite{Ner11,Tim11,Rid12,Rid13_2,Kir13,Bul13}. For brevity we will describe the resulting state as a 'QED plasma'. 

\begin{figure}[h!]
\centering	
\includegraphics[width=0.4\textwidth]{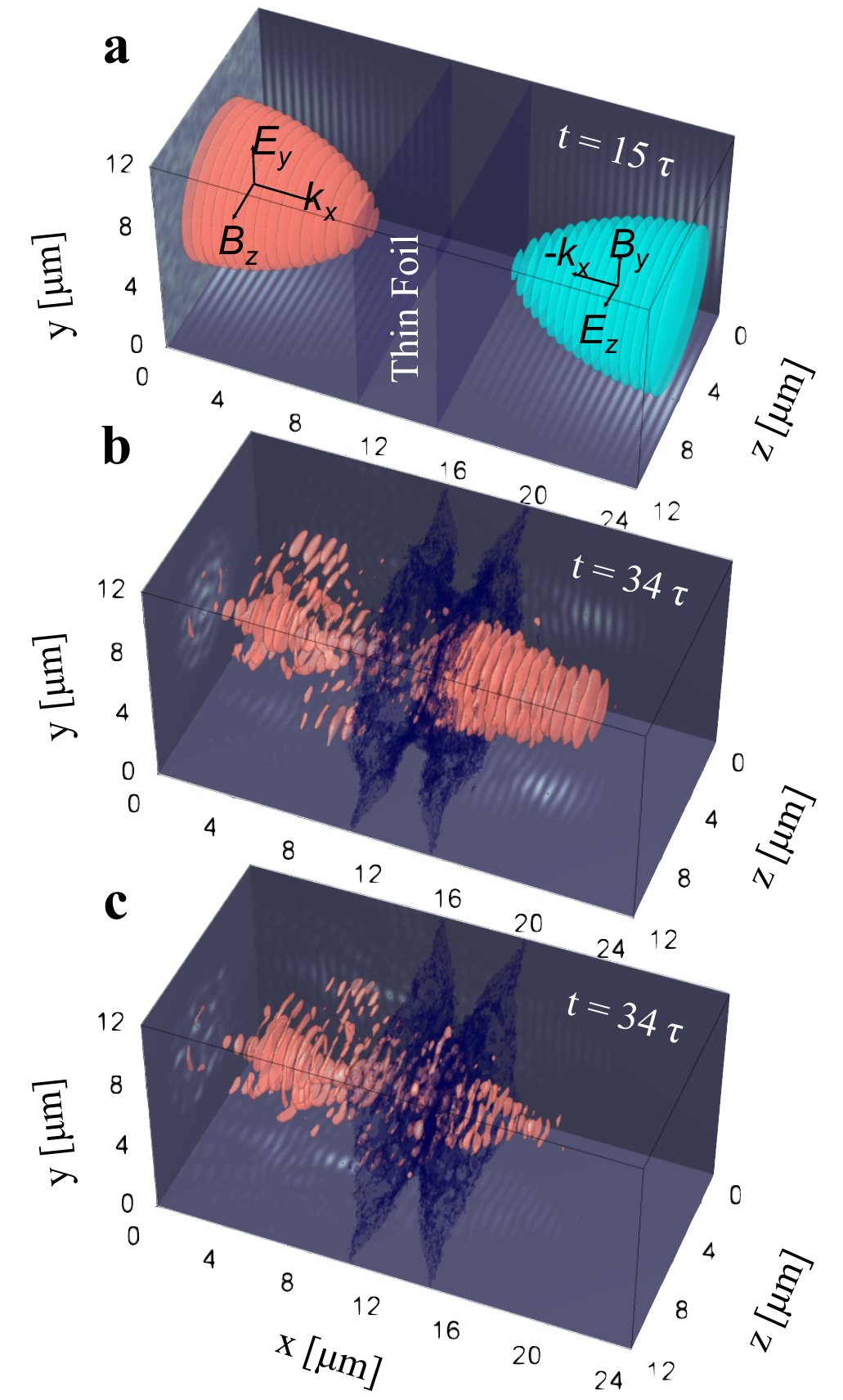}
\caption{\label{fig:fig1} 
3D QED-PIC simulation (a), Two laser pulses illuminate a thin foil from both sides. The left pulse is linearly polarized in the $y$-direction, and the right pulse is linearly polarized in the $z$-direction. (b-c), the magnitude of electric field $E_y$ and the electron density profile at $t = 34 \tau$, for the QED-off and QED-on simulations, respectively. Side panels display slices of $E_{y(z)}^2$ through the center of the box.
}
\end{figure}

In non-relativistic plasma, electromagnetic waves with frequency $\omega_0<\omega_{pe}=(n_ee^2/m_e\epsilon_0)^{1/2}$ cannot propagate, where $n_e$ is the electron number density.  The critical density $n_c$ is the density at which the plasma frequency equals the wave frequency and above which the plasma is described as `overdense'.  In the ultra-relativistic regime,  electrons in the plasma are accelerated by the laser fields to such high energy that they have an effective mass much greater than their rest mass. Consequently the plasma frequency is  reduced by a factor $1/\sqrt{\langle\gamma\rangle}$, where $\langle\gamma\rangle$ is the average Lorentz factor of the electrons.  An opaque (and nominally overdense) plasma may therefore be expected to become transmissive if the laser intensity, and therefore $\langle\gamma\rangle$, is sufficiently high. This ``relativistically induced'' transparency \cite{Kaw70,Pal12} optically switches the plasma from opaque to transparent and enables light propagation.   However, as the electromagnetic fields increase and we enter the QED-plasma regime, radiation reaction becomes significant, the electron motion is damped and hence $\langle\gamma\rangle$  reduced.   Furthermore, radiation-reaction  leads to absorption of the electromagnetic wave \cite{Bas13}.  At higher intensities still, pair plasma may be produced that is sufficiently dense to shield the laser fields.  As a result of these processes, a classical prediction that a plasma is transmissive can be erroneous when QED effects are introduced.

Consider a potential experiment with two counter propagating laser pulses of intensity $I>10^{22}\;{\rm Wcm^{-2}}$ with orthogonal linear polarizations impinging normally on both surfaces of a solid density foil. The radiation can be measured after the interaction with a calorimeter to determine the energy of the pulse after the interaction and because of the orthogonal polarizations it can be determined what radiation is reflected or transmitted. In this Letter we will show that, for an appropriate density/thickness foil, the laser radiation should be  efficiently absorbed provided strong-field QED effects are considered in the plasma model, but if they are neglected the foil will be completely transmissive. 

To model this experiment, we performed large-scale three-dimensional (3D) numerical simulations using the QED-particle-in-cell (QED-PIC) code EPOCH {\cite{Bra11,Rid13,Kir09,Sok09,Duc11,DiP10}},  
as shown in Fig. \ref{fig:fig1}. EPOCH extends the Vlasov-Maxwell system to include the important QED processes in next generation laser-plasma interactions and is detailed in Ref. \cite{Rid13}.  The electromagnetic field is split into high and low frequency components. The low frequency components are coherent states that are unchanged in QED interactions \cite{Gla63}.  The evolution of these macroscopic fields is determined by solving Maxwell's equations.  Electron and positron basis states are `dressed' by these low-frequency fields, which are treated as a classical background in the interactions of these charged particles with the high frequency component of the field, i.e. using the strong-field QED or `Furry' representation \cite{Fur51}.  We include the dominant first-order processes: gamma-ray photon emission by electrons and positrons and pair production by gamma-ray photons.  Electron and positron motion between and during these interactions is described by the `quasi-classical' model of Baier and Katkov \cite{Bai68}.

Simulations were performed for  two conditions: (1) "QED-off", by which we mean no gamma-ray photons or pairs are generated and (2) "QED-on", by which we mean that the QED-PIC code produces gamma-ray photons, tracks the photon's dynamics and generates electron-positron pairs.  
The target is 4~$\mu$m thick, with an initial density of 150~$n_c$. The ion mass is taken as $m_i/m_e = 3674$ and the ion charge as $q/e = 1$. The lasers propagate in the $\pm{x}$ direction, as shown in Fig. \ref{fig:fig1}a. The left-hand-side laser pulse is polarized in $y$ direction and the right-hand-side laser pulse is polarized in $z$ direction. Both the laser pulses are of intensity $I = 4.5\times{10}^{23}$Wcm$^{-2}$, which is within the range expected to be reached by next-generation 10 PW lasers, with a gaussian profile of FWHM radius 4~$\mu$m. Both laser pulses are  $25$ fs in duration. The radiation pressures of the laser pulses from opposite sides balance, such that the target is expected to be confined and centered around its initial position. However, it is compressed until the thermal pressure balances the ponderomotive pressure.  

 After the laser pulses reached the surfaces of target foil, the foil was compressed inwards from both sides. For the QED-off simulation, when the peaks of the laser pulses arrived, both pulses started penetrating through the foil and eventually were transmitted. However, for the QED-on simulation, the laser pulses were blocked by the foil during the entire course of the simulation, indicating that relativistic transparency was suppressed by the QED effects. The magnitude of electric field $E_y$ and the electron density profile are plotted at a time of $34\tau$, where $\tau$ is the laser period, in Figs. \ref{fig:fig1}b and c, for the QED-off and QED-on simulations, respectively. 

\begin{figure*}
\centering
	\includegraphics[width=\textwidth]{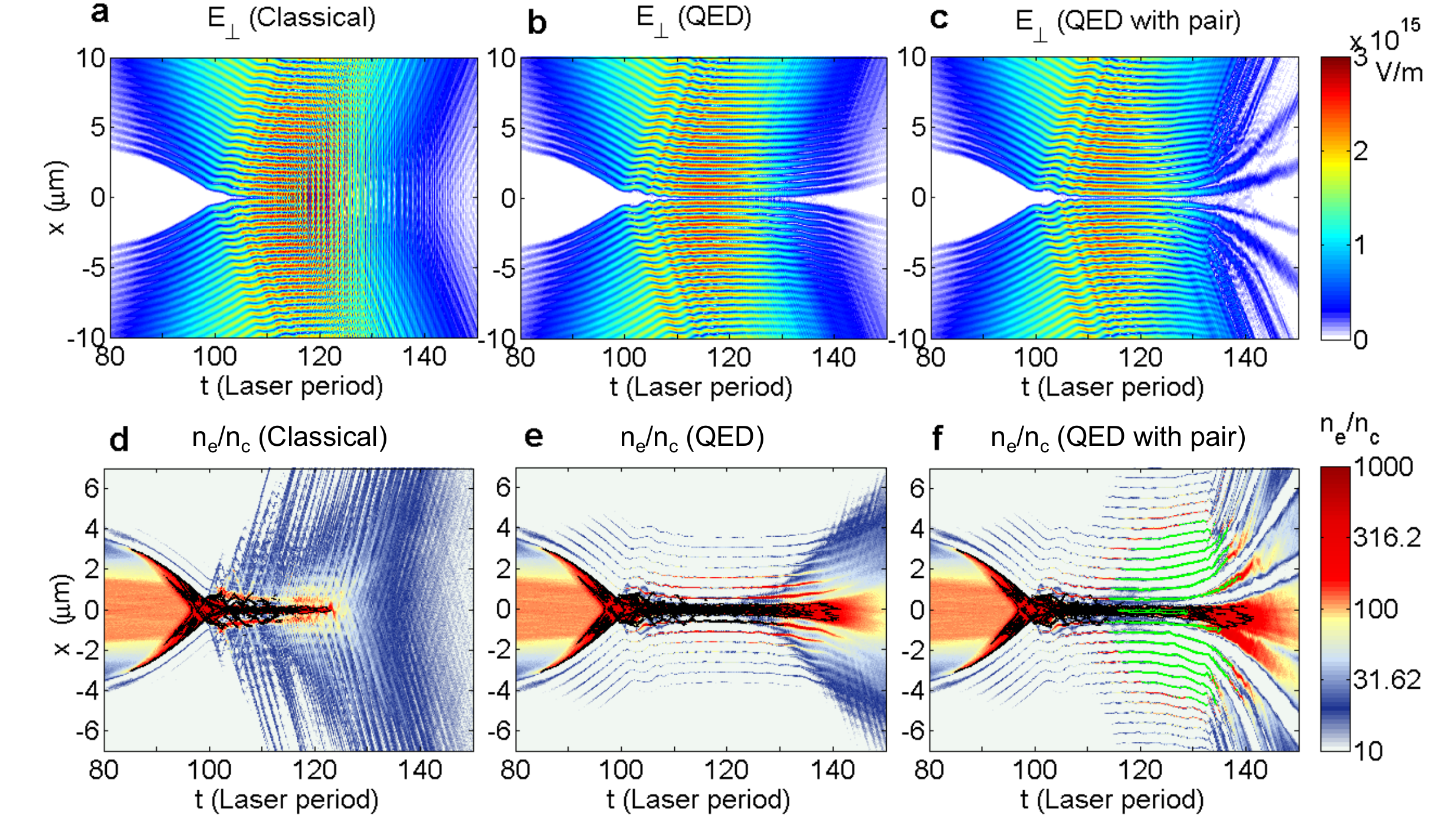}
\caption{\label{fig:fig2}  
Magnitude of laser field $E_{\perp}$ and electron density $n_e/n_c$ as a function of time $t$ and position $x$ from 1D QPIC simulation. (a)-(c) Magnitude of laser field $E_{\perp}$, (d)-(f) electron density $n_e/n_c$. Also plotted are the contours for ion density $n_i/n_c$ (black lines) and positron density $n_p/n_c$ (green lines).(a) and (d) QED is switched off, (b) and (e) QED is switched on but with photon emission only, so that the (quantum corrected) radiation reaction effects are included, and (c) and (f) QED is fully activated to include both photon emission and pair production.
}
\end{figure*}

To facilitate comparison to analytical theory, we repeated the above simulation with circularly-polarized laser pulses, using 1D QED-PIC code.  Very similar results on the transmission behavior were obtained  as compared to the above 3D case. This is because the linear polarizations are similar to circular polarization at particular points in the overlap between the two lasers, in particular in the center of the domain.   The laser pulse illuminating the target at normal incidence has a wavelength of $1$~$\mu$m and a gaussian profile with a  pulse duration of 60 fs and with intensity $I = 5\times{10}^{23}$ Wcm$^{-2}$. The initial electron density is $n_0 = 100n_c$, where $n_c = \omega_0^2m\epsilon_0/e^2$ is the non-relativistic plasma critical density. The target is a slab of thickness  4~$\mu$m; the ion mass is taken to be $m_i/m_e = 3674$ and the ion charge as $q/e = 1$.  The simulations were performed under three conditions: 1) QED-off; 2) QED-on with gamma-ray photon generation only (the code is allowed to produce photons and calculates the recoil due to emitting photons, which gives the radiation reaction effect, but the photon is not tracked and thus there is no pair production involved); and 3) QED-on with both gamma-ray photon and pair production.

The magnitude of the laser electric field and the plasma density are plotted as a function of time $t$ and position $x$, as shown in Fig. \ref{fig:fig2}, for all the three conditions. For all the three conditions, the laser plasma interaction process may be characterized by three distinct stages: an initial piston-like forward push of electrons by the laser illumination, followed by stagnation at the center between the two lasers and finally electron re-injection in the backward direction. 

During the initial push, both electrons and ions are continuously compressed inwards until $105\tau$, as shown in Figs. \ref{fig:fig2} (d-f). This compression process is almost identical under all the three conditions, implying that radiation reaction and QED effects are not important at this stage. 

In the case of a single laser beam hitting a charged particle at rest, the zero momentum frame of reference (ZMF), in which the charged particle has a periodic trajectory, does not coincide with the laboratory frame due to particle recoil. ZMF moves in the direction of propagation of the laser beam with a velocity corresponding to a Lorentz factor equal to $a$ {\cite{Bel08,Lau03}}. The laser frequency in the ZMF is thus redshifted compared to the laboratory frame. Because the strength parameter $a$ is a Lorentz invariant, the reduced wave frequency in the ZMF implies a reduced amplitude of the wave field: $E_{RF} \approx E/a$ and so $\chi\ll1$ and the effects of radiation reaction and QED become insignificant. 

After the piston-like push of the electrons by laser, the bulk plasma stagnates when the ponderomotive pressure of the laser is balanced by the thermal pressure of the compressed plasma slab. We find that in this quasi-static phase of the two sided illumination only the electron motions need to be considered since there is no bulk ion motion.

There was always a small portion of the electrons at the edge of the electron layer moving back into the laser fields, regardless whether QED  effects are included or not. This is simply because there is no net longitudinal force on a thermal electron.  However, radiation reaction dramatically changes the dynamics of the back-injected electrons to radiatively cool them. They form equally spaced ultra-high-density thin electron layers in the nodes of the standing wave formed by the incident and reflected wave, as shown in Figs. \ref{fig:fig2}e and f. These cause the plasma to become optically opaque compared to the QED-off case where the layers are not present and the plasma is transmissive. This is shown in Figs. \ref{fig:fig2} (a-c). 

The situation of  electrons circulating in an electric field in the presence of a fixed ion background  --- which is similar to the stagnated layer discussed previously since no bulk ion motion is involved --- readily lends itself to analytical solution.  We have confirmed this using 1D QED-PIC simulations with fixed ion background. The force balance in the longitudinal direction (between the electrostatic force and light pressure) means that to good approximation the electron's circulate in the transverse direction under the influence of the transverse laser fields.  Thus the more straightforward, `zero-dimensional', solution for a wave propagating in a homogeneous plasma is of relevance.  An equilibrium solution can be found for the wave equation for the laser, ${\bf A}^{\prime\prime}-c^2\nabla^2{\bf A} = \rho_0c({\boldsymbol\beta}_i-{\boldsymbol\beta}_e)/\epsilon_0$, with the electrons and ions co-rotating with constant $\gamma$ transverse to the propagation. The electron current contribution is larger by a factor of approximately $m_i/(m_ea_0)$, so for simplicity we neglect the ions.
  
Classically, the electron equation of motion for an electron in a circulating electric field, including radiation reaction according to the Landau-Lifshitz prescription \cite{Lan75,Bel08}, is
\begin{equation}
\frac{d(\gamma\boldsymbol{\beta})}{dt}=\frac{eE}{m_{e}c}\left\{\hat{\bf e}-\frac{2e^{3}}{3m_e^{2}c^4}\gamma^2\boldsymbol{\beta}E\left[1-(\boldsymbol{\beta}\cdot\hat{\bf e})^2\right]\right\}
\label{LLs}
\end{equation}
where $\hat{\bf e}$ is the unit vector in the direction of the electric field. Although we are considering an electro\emph{magnetic} wave propagation, near the threshold for propagation the magnetic field  will be much smaller than  the electric field. It will also be close to parallel to the electron momentum for all but the strongest radiation damping. It is therefore reasonable to neglect its contribution to the electron motion and it greatly simplifies the analysis.

\begin{figure}
\centering
	\includegraphics[width=0.5\textwidth]{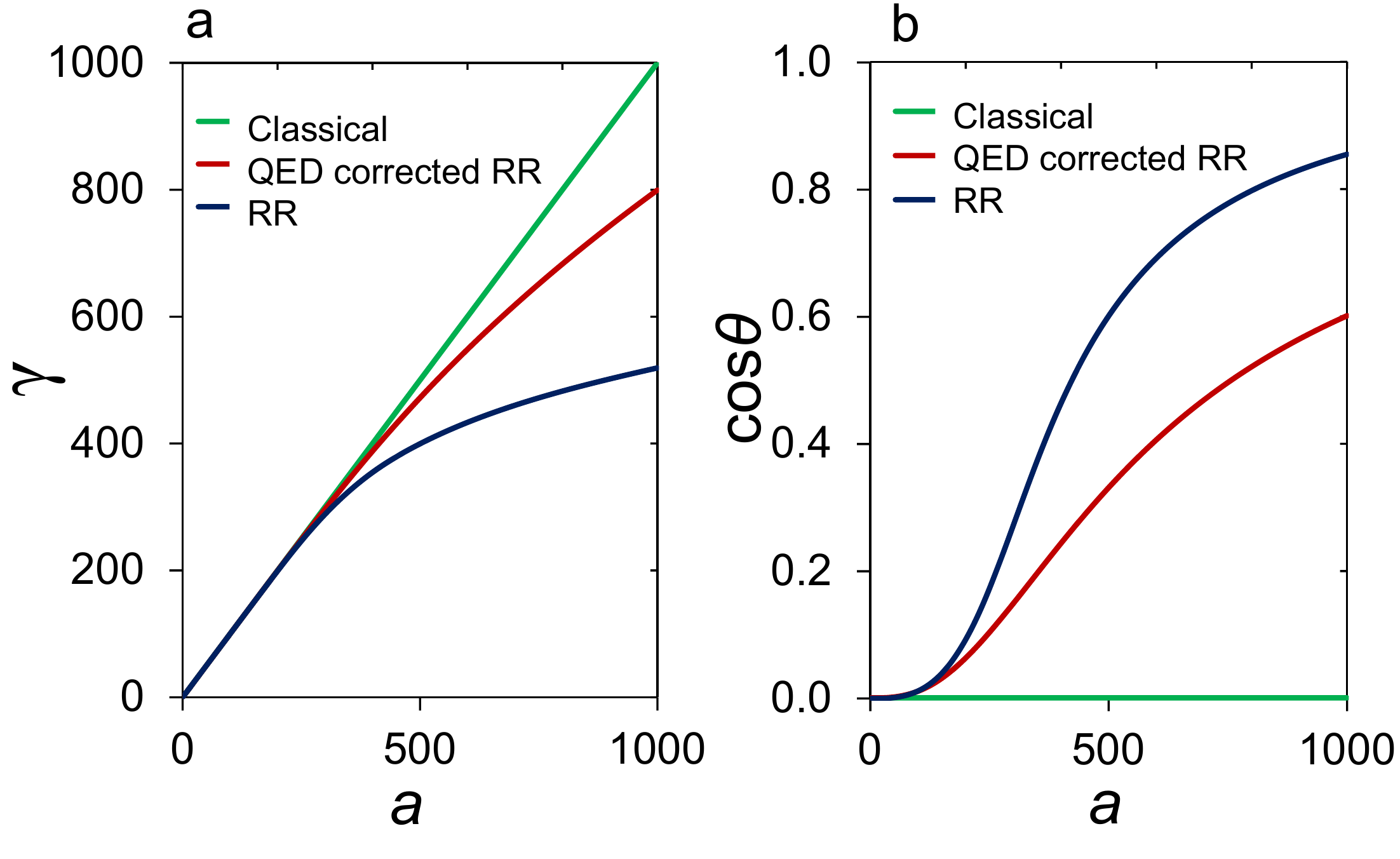}
\caption{\label{fig:fig3} 
Effects of radiation reaction on single electron dynamics. (a) $\gamma$ and (b) $\cos\theta$ as a function of $a$ for $\lambda=1\mu{m}$, under three conditions: classical, i.e. no radiation reaction force, $\gamma = (1+a^2)^{1/2}$; RR, with classical radiation reaction force, Eqn. (\ref{LLsga}); and QED corrected RR, with quantum corrected radiation reaction force.
}

\end{figure}

An equilibrium solution to Eq. (\ref{LLs}) exists where the
component of electric field parallel to $\boldsymbol{\beta}$ precisely compensates for the radiative losses, while the perpendicular component
 provides the centripetal force for  circular motion of the electrons. Setting the LHS of Eq. (\ref{LLs}) to 0, we find, by neglecting terms of order $\gamma^{-2}$ or smaller, 
\begin{equation}
\gamma=a\sin\theta,\;\;
\Gamma\sin^2\theta-\cos\theta=0,
\label{LLs2}
\end{equation}
where $\Gamma = (2/3)(e^2/m_ec^3)\omega_0a\gamma^2$. Combining Eq.
(\ref{LLs2}) and eliminating $\theta$, we obtain a relation between $\gamma$ and $a$ 
\begin{equation}
\tau_R^2\omega_{0}^2\gamma^{8}+\gamma^{2}=a^{2}
\label{LLsga}
\end{equation}
where $\tau_R=2e^2/3m_{0}c^3=4.163\times{10}^{-24}$~s. Equation (\ref{LLsga}) is plotted in Fig. \ref{fig:fig3}a for $\lambda=1\mu$m. For $a < 300$, $\gamma\cong$$a$, radiation reaction effects on the dynamics of the electrons are negligible. For $a > 300$, the value of $\gamma$ becomes smaller than $a$, as  electron motions are inhibited by the strong radiation reaction force.

It is well known \cite{Kir09,Erb66,Rit79,Tho12} that the classical description of an electron radiating in a strong electromagnetic
field overestimates the total emitted power. This understanding is connected to the fact that, in the quantum description, the emitted photon energy may not exceed the electron energy, whereas the classical approach does not have such a restriction. This effect may be approximately taken into account as follows. The total power of emitted radiation can be expressed as the power calculated from the classical description of radiation reaction, times a reducing factor $g(\eta)$ that accounts for truncation of the emitted spectrum \cite{Kir09,Rit79,Tho12}. 
Eqn. (\ref{LLs2}) still holds, except that now $\Gamma = g(\eta)(2/3)(e^2/m_ec^3)\omega_0a\gamma^2$. $\gamma$ with quantum corrected radiation reaction effect is also plotted in Fig.\ref{fig:fig3}a. In general, quantum effects reduce the classical radiation back reaction, which make the Lorentz factor $\gamma$ of the electrons only slightly smaller than its conventional value $\gamma \cong a$. However, the fraction of the laser power that produces synchrotron radiation, which is proportional to the electric field parallel to electron velocity, i.e. $\cos\theta$, remains significant, as shown in Fig.\ref{fig:fig3}b. 

The significance of this can be seen by considering the resulting wave equation, which including this current contribution yields a dispersion relation 
 \begin{equation}
\omega^2 = k^2c^2 +\frac{\omega_{p0}^2}{\gamma\left(1-i\tau_R\omega_0\gamma^3\right)}\;,
\label{dispersion}
\end{equation}
that for $\tau_R\omega\gamma^3\ll1$ has a real component $\omega_0 = \sqrt{k^2c^2+\omega_{p0}^2/\gamma}$ and an imaginary component, i.e. representing \emph{absorption} of the wave, of $\omega_I = \omega_{p0}^2\tau_R\gamma^2/2$, with $\gamma$ given by Eq. (\ref{LLsga}). Since $\gamma\simeq a$, $\omega_0\simeq\sqrt{k^2c^2+\omega_{p0}^2/a}$ and $\omega_I\simeq \omega_{p0}^2\tau_Ra^2/2$. This gives a critical frequency for the threshold for wave propagation, i.e. where $k=0$, of $\omega_c = \omega_{p_0}/\sqrt{a}$. 

 In the strong damping limit, $\tau_R\omega\gamma^3\gg1$, the dispersion relation yields an imaginary component $\omega_I = \omega_{p0}^2/(2\omega_0^2\tau_R\gamma^4)$ and for $\omega_{p0}^2/(\gamma\omega_0^2)\gg1$, $\omega_0^2 = {k^2c^2+\omega_{p0}^4/(4\omega_0^4\tau_R^2\gamma^8)}$. In this limit, $\gamma\simeq \left[a/(\omega_0\tau_R)\right]^{1/4}$ so that $\omega_I =  \omega_{p0}^2/(2\omega_0a)$. The critical frequency for wave propagation is now $\omega_c=\omega_{p_0}/\sqrt{2a}$, which is a factor of $\sqrt{2}$ smaller than that of weak damping limit. However, at this frequency the wave is critically damped, i.e. $\omega_I = \omega_c$, and will therefore be attenuated rapidly. 

In summary, we have studied the interplay of QED effects and collective processes in plasmas generated by next-generation 10~PW lasers.  In particular we have derived a dispersion relation for electromagnetic wave propagation and shown that the damping of collective electron oscillations leads to strong absorption of the wave.  This has dramatic consequences for relativistic transparency. QED-PIC simulations are performed to demonstrate that the relativistically transparent plasma may be converted to become optically opaque if  QED effects become important. 3D simulations were performed to provide experimental framework to test the relativistic transparency with strong radiation reaction and QED effects and the scaling of the absorption/transmission with laser intensity should be reasonably easy to measure experimentally. 
These
predictions may be tested using high-power lasers in the next few years. 

This research was supported by the AFOSR and in part through computational resources and services provided by Advanced Research Computing at the University of Michigan, Ann Arbor.


\end{document}